\documentclass{pinchcr}
\usepackage{amsmath}
\usepackage{amsfonts}
\usepackage{amssymb}

\title{Competing instabilities in quench experiments with ultracold Fermi gases near a Feshbach resonance}

\author{David Pekker}
\affiliation{California Institute of Technology, Pasadena, CA 91125, USA}

\author{Eugene Demler}
\affiliation{Harvard University, Cambridge, MA 02138, USA}

\authors{2}

\begin{document}
\maketitle
\maintext

\section{abstract}
Tunability of effective two body interactions near Feshbach
resonances is a powerful experimental tool in systems of ultracold atoms. 
It has been used to explore a variety of intriguing phenomena in recent experiments. 
However not all of the many-body properties of such systems
can be understood in terms of effective models with contact interaction
given by the scattering length of the two particles in
vacuum. For example, when a two component Fermi mixture is quenched to
the BEC side of the Feshbach resonance, a positive scattering length
suggests that interactions are repulsive and thus collective dynamics should be dominated by the 
Stoner instability toward a spin polarized ferromagnetic state. On the other hand,
existence of low energy two particle bound states suggests a competing
instability driven by molecule formation. Competition between
spontaneous magnetization and pair formation
is determined by the the interplay of two-particle and many-body
phenomena. In these lecture notes we summarize our recent theoretical
results, which analyzed this competition from the point of view of
unstable collective modes. We also comment on the relevance of this
theoretical analysis to recent experiments reported in Ref.~\cite{Ketterle}. 

\section{Introduction}
It is often effective to characterize many-body systems in terms of their emergent collective modes that describe their low energy excitations. Examples of collective modes include sound waves in interacting gases, spin (magnetization) waves in magnetic systems, and phase modes in superconducting systems. The strength of the collective modes approach is that for many cases, such as Fermi liquids, ferromagnets, superconductors, etc. these collective modes are long lived and weakly interacting. Moreover, collective modes can be useful for understanding not only properties of systems near equilibrium but also for the dynamics of systems away from equilibrium.

The purpose of these lecture notes is to discuss how we can utilize the analysis of unstable collective modes to study the dynamics of fermionic systems quenched from the weakly interacting state to the strongly interacting regime in the vicinity of the Feshbach resonance. Conceptually this approach is similar to the spinodal decomposition in statistical physics~\cite{Chaikin_Lubemsky_textbook} and analysis of domain formation in unstable bosonic systems, such as immiscible  two component mixtures~\cite{StamperKurnKetterle}, quenched ferromagnetic spinor Bose condensates~\cite{StamperKurn,Lamacraft}, dynamics of spiral states in ferromagnets~\cite{Cherng,ConduitAltman}. The main new feature of the fermionic problem that we consider is the need to provide careful regularization of the strong two-body interactions arising from Feshbach resonances. As we show the interplay of strong two-body interactions and many-body effects, such as a Pauli blocking by the Fermi sea, lead to some very intriguing aspects of dynamics. 

Non-equilibrium dynamics of quantum systems has been studied for a long time. However, until recently, experimental ability to control quantum systems has been rather limited. The progress in parametric control of quantum Hamiltonians of ultracold atom systems has brought renewed interest to the field of non-equilibrium dynamics. One such recent experiment, performed by the MIT group, studied ramps of a two component Fermi gas from the weakly repulsive to the strongly repulsive regime~\cite{Ketterle}. The ramp was accomplished by sweeping the magnetic field to a value close to the Feshbach resonance, but on the BEC side. Strongly repulsive fermions are known to be susceptible to the Stoner instability to ferromagnetism~\cite{Stoner}. In fact, this very instability is believed to be responsible for itinerant ferromagnetism in metals. Intriguingly, the MIT experiment observed several surprising phenomena, which they interpreted as signatures of ferromagnetism, yet they did not find any ferromagnetic domains. However, strongly repulsive ultracold fermions come with a price --- the repulsion is a byproduct of a shallow bound state. Hence, there is an alternative explanation for the MIT observations --- susceptibility to pair formation~\cite{us}. In these notes our goal is to understand many-body fermionic systems during and immediately after a quench into a strongly interacting regime near a Feshbach resonance. Emphasis of our discussion will be on understanding the interplay of two-particle and many-body aspects of dynamics.

Our study can be placed in a general perspective of quenches that take a quantum system from a disordered phase to an ordered phase described by some order parameter (e.g. magnetization or pairing). We show that the dynamics is initially driven by local instabilities towards ordering. A useful way to understand these instabilities is in terms of collective modes. In equilibrium, all collective modes of a system are, by definition, stable. Thus the dynamics following a small perturbations of a collective coordinate is either periodic or damped but not growing. In other words, in equilibrium the imaginary part of the collective mode frequency is either zero or negative (we adopt the convention that $O(t) \sim O_\omega e^{ - i \omega t}$). On the other hand, consider the dynamics of a system following the parametric ramp or a quench across a phase transition.  In order to come to equilibrium on the ordered side, the system must develop a finite order parameter following the quench. Generically, following the quench, some of the collective modes of the system become unstable, that is they acquire a positive imaginary frequency. This local collective mode instability corresponds to the growth of the order parameter, and leads to its ``nucleation". For the case of the MIT system, the quench leads to two distinct types of incompatible instabilities: pair formation and ferromagnetism. The competition between the two ultimately determines the fate of the quantum system. 

Although a complete description of dynamic processes following a ramp across a phase transition remains to be found, we will present some aspects of the problem. The problem of understanding dynamics of large quantum systems is that there are no generic methods of attack. Therefore, one may want to start by considering small systems that are susceptible to methods like exact diagonalization. In small quantum systems there are no true phase transitions. However, one can think about parametric tuning of the Hamiltonian that leads to a level crossing associated with the phase transition of the thermodynamic system. For a finite system, such level crossings are typically associated with the changing of an approximate symmetry of the ground state and therefore they are avoided level crossing. The dynamics of the quantum system are then described by the Landau-Zener process~\cite{LZ1,LZ2}, depending on the ramp rate the system may either remain in the ground state or it may jumps to the excited state. If we try to scale up these arguments to the case of a large many-body system we immediately run into a problem. During a parametric ramp, the exponentially large number of eigenstates of the many-body system undergo ``spectral-flow": some of these eigenstates move up and some down. The result of the spectral-flow is a large number of Landau-Zener process, which generically require an exponentially large Hilbert space to describe. Therefore, a method that would keep only the ``important" excitations is very useful.

The approach of studying the collective modes in order to describe the dynamics following a parametric ramp or quench has a long history. Kibble~\cite{Kibble} and Zurek~\cite{Zurek} were the first to point out that a short time after a thermal quench to an ordered state (e.g. superconducting state), distant parts of the systems could not have exchanged information and therefore the order parameter at large distances cannot be correlated. Hence, shortly after a thermal quench it is possible to find locations in space around which the order parameter winds, thus the dynamics following a quench results in the formation of topological defects (e.g. vortices for the superconducting case). Further dynamics involves the motion and recombination of the topological defects produced at short time scales. The specific problem of the motion and recombination of topological defects has been studied extensively for the case of thermodynamic phase transitions in various systems including liquid crystals. Liquid crystals, in particular, have attracted attention due to both their technological application in displays and the relative accessibility of their dynamics from the experimental perspective as the collective modes and topological defects associated with the various ordered states can be observed using polarized light~\cite{LC}. The dynamics of these systems following a quench conform to the notion of initial growth followed by power-law slow defect recombination~\cite{Bray}. 

To summarize, the dynamics can be split into two time scales. An initially short time-scale during which the various unstable collective modes grow exponentially. As these modes are largely non-interacting, they can be treated as independent modes in momentum space. As the amplitudes of the unstable collective modes grow, they start to interact with each other. When the interactions become strong, the resulting order parameter pattern produced by the initial growth ``freezes-in." Consequently, the dynamics is controlled by the motion of topological defects (as well as non-topological excitations). 

In these lecture notes, we shall apply the method of collective mode instabilities to the concrete example of quenches of the non-interacting Fermi gas to the strongly interacting regime.  The motivations for this are (1) The existence of ferromagnetism in an itinerant fermionic system is a long standing open question in condensed matter physics. While ferromagnetic transition can be predicted based on a  simple mean-field analysis\cite{Stoner} several objections to this argument can be raised. Kanamori was the first to point out that screening was essential to understanding the Stoner transition, and may even prevent it~\cite{Kanamori}. For example, the Stoner criterion suggests ferromagnetic instability even in cases when rigorous theorems forbid such a transition, such as one dimensional systems~\cite{Lieb-Mattis}. Recent theoretical investigations have found that the transition survives in 3d but becomes of first order type at low temperatures~\cite{StonerTransition1,StonerTransition2,StonerTransition3,StonerTransition4,StonerTransition5,StonerTransition6,StonerTransition7}. Therefore, if results of the MIT experiments can indeed by interpreted as Stoner type instability, then they resolve a fundamental matter of principle question in physics. However before we accept this interpretation, it is important that we rule out competing scenarios, such as the nonequilibrium dynamics of pairing. Other arguments against the possibility of observing Stoner instability in ultracold Fermi gases near Feshbach resonance have been given in Refs.~\cite{Zhai,Ho,Zwerger}
(2) The true ground state in such system is a condensate of molecules. It was proposed that a Stoner ferromagnetic state can be created dynamically  if the rate at which magnetic correlations develop is considerably faster than the rate of molecule formation~\cite{Ketterle,Pilati,Trivedi}. Whether such hierarchy of instabilities really takes place is crucial for interpreting experimental results. (3) For this reason, studying the dynamics following a quench is relevant to the current experiments. (4) Studying the Stoner and pairing instabilities is a good way to discuss Feshbach resonance in many-body systems. Throughout we shall discuss the pairing and the Stoner instabilities from the perspective of a geometrical Feshbach resonance (in which the quantum numbers of fermion in the molecular state are identical to those of free fermion). (5) Our analysis has important general implications for the idea of quantum simulations with ultracold atoms. In many cases one is interested in using strong repulsive interactions between atoms in order to create analogues of condensed matter systems.  Our analysis provides a warning that to get any meaningful results, one needs to make sure that molecule formation does not overwhelm dynamics determined by the repulsive interactions.

The notes are organized as follows. In Section~\ref{sec:LR} we begin by discussing the relation between linear response and collective modes. We also show how to compute the pairing and the ferromagnetic responses using the equation of motion formalism. Next, in Section~\ref{sec:PP}, we show how to describe a Feshbach resonance using a pseuodpotential model. In Section~\ref{sec:Apply}, we apply the pseudopotential model to compute the many-body T-matrix and thus obtain the pairing collective mode. We comment on how to incorporate the many body T-matrix into the ferromagnetic susceptibility in Section~\ref{sec:StonerWithCooperon}. Finally, in Section~\ref{sec:Discussion} we summarize the results for the pairing vs. Stoner competition in the context of the MIT experiments, and make concluding remarks in Section~\ref{sec:Conclude}.

\section{Linear response and collective modes}
\label{sec:LR}
\subsection{From poles of the response functions to collective modes}
A useful approach to think about collective modes is via their link to linear response susceptibilities. A linear response susceptibility $\chi_A(q,\omega)$ is the defined as the link response of some property of the system $\langle A(q,\omega) \rangle$ under the influence of an external perturbation of strength $h_A(q,\omega)$ that is thermodynamically conjugate to it
\begin{align}
\langle A(q,\omega) \rangle = \chi_A(q,\omega) h_A(q,\omega).
\end{align}
Here, by thermodynamically conjugate, we mean that the external perturbation $h_A(q,\omega)$ contributes a term to the Hamiltonian
\begin{align}
{\mathcal H}_\text{external}=h_A(q,\omega) e^{i \omega t} \hat{A}(-q,\omega) + h. c..
\end{align}
Generally, to obtain a finite value of $\langle A(q,\omega) \rangle$ one needs to have a non-zero value of the external field $h_A(q,\omega)$. An important exception is when the response function $\chi_A(q,\omega)$ has a pole and is therefore infinite. Writing the integral equation for the response of the system
\begin{align}
\langle A(q,t)\rangle =\int dt' \, \chi_A(t-t',\omega) h_A(q,t'),
\end{align}
we see that poles of $\chi_A(q,\omega)$ correspond to long lived modes of the system, which are identified as the collective modes~\cite{AGD,NozieresPines}. The argument above is conventionally used for systems with time independent parameters. In the next section we will discuss how it can be extended to the case when the interaction strength is changing in time. 

\subsection{Dynamics of the pairing amplitude using time-dependent Hartree approximation}
One approach to obtain a response function and thus the associated collective mode spectrum is by using the equation of motion formalism for the operator $\hat{A}(q,t)$. We shall now follow this procedure in detail for the case of the pairing mode of a two component Fermi gas near a Feshbach resonance. In doing so we shall demonstrate the RPA approximation that is often used to brining the equation of motion $\dot{A}=i\,[H,A]$ into a closed form. 

Consider an interacting gas composed of two species of fermions. Suppose that the gas may be described by the Hamiltonian
\begin{align}
H=\sum_{k,\sigma} \epsilon_{k,\sigma} c^\dagger_{k,\sigma} c_{k,\sigma} + \sum_{q} \hat{\rho}_{q,\uparrow} V_{q} (t) \, \hat{\rho}_{q, \downarrow},
\end{align}
where, $c^\dagger_{k,\sigma}$ and $c_{k,\sigma}$ are the creation and annihilation operators for a fermion of species $\sigma$ and momentum $k$ and non-interacting energy $\epsilon_{k,\sigma}=k^2/2m_\sigma-\mu$. $V_q(t)$ describes the time dependent inter-atomic potential undergoing parametric tuning (we shall assume s-wave scattering) and $\hat{\rho}(q,\sigma)=\sum_k c^\dagger_{k+q,\sigma} c_{k,\sigma}$ is the density operator. To probe the pairing susceptibility, we add the thermodynamically conjugate external perturbation 
\begin{align}
H_\text{ext}={\mathcal P}_q^\text{ext}(t) \sum_k c^\dagger_{q/2+k, \uparrow}(t) c^\dagger_{q/2-k, \downarrow} (t) + h.c.,
\end{align}
and measure the pairing amplitude ${\mathcal P}_q(t)=\sum_k \langle c^\dagger_{q/2+k, \uparrow}(t) c^\dagger_{q/2-k, \downarrow}(t)\rangle$. 

At this point, it is useful to introduce the operator $B(k,q,t)=c^\dagger_{q/2+k, \uparrow}(t) c^\dagger_{q/2-k, \downarrow}(t)$ which is related ${\mathcal P}_q(t)$ by a summation over $k$ and taking the expectation value. The equation of motion for the operator $B$ is
\begin{align}
\frac{d}{dt} B(k,q,t) = i [H, B(k,q,t)].
\end{align}
Evaluating the various commutators we obtain
\begin{align}
i& \frac{d}{dt} B(k,q,t) = (\epsilon_ {q/2+k, \uparrow}+\epsilon_{q/2-k, \downarrow}) B(k,q,t)  \nonumber \\
&
+ [1-\hat{n}_\uparrow(q/2+k,t)-\hat{n}_\downarrow(q/2-k,t)]\left[{\mathcal P}_q^\text{ext}(t) + \int {d\!\!^{-}\!} p \, V_{k-p}(t) B(p,q,t) \right], \label{eq:EOM}
\end{align}
where $\hat{n}_\sigma(q,t)= c^\dagger_{q, \sigma}(t) c_{q, \sigma}(t)$ is the number operator, and ${d\!\!^{-}\!} p$ stands for $d^3p/(2\pi)^3$. By taking the expectation value of Eq.~\eqref{eq:EOM}, we find that the equation of motion for the two fermion expectation value $\langle B(k,q,t) \rangle$ will be coupled to expectation values containing four fermion operators. This coupling to higher order expectation values is a general feature of equation of motions for interacting theories. To obtain a closed form equation, we must cut off the equation of motion at some point. A typical approach is called the Random Phase Approximation (RPA), which states that the paring amplitude must oscillate at the drive frequency, and therefore one can decouple the pairing amplitude in four fermion terms. That is, within the RPA we replace the four fermion expectation values by the product of the two fermion ones: $\langle \hat{n}_\uparrow(q/2+k,t) B(p,q,t)  \rangle \rightarrow \langle \hat{n}_\uparrow(q/2+k,t) \rangle \langle B(p,q,t) \rangle$. Further, as  the expectation value $\langle \hat{n}_\uparrow(q/2+k,t) \rangle$ is presumed to be stationary by the RPA, we can replace it by its value in the initial state $n^F_\uparrow(q/2+k)$. The RPA assumption is consistent with the assumption that only the unstable collective mode has interesting dynamics, and therefore we can ignore the dynamics of other expectation values. Taking the expectation value of the equation of motion and making the RPA approximation we obtain
\begin{align}
i &\frac{d}{dt} \langle B(k,q,t) \rangle = (\epsilon_ {q/2+k, \uparrow}+\epsilon_{q/2-k, \downarrow}) \langle B(k,q,t) \rangle \nonumber \\
&
+ [1-n^F_\uparrow(q/2+k)-n^F_\downarrow(q/2-k)]\left[{\mathcal P}_q^\text{ext}(t) + \int {d\!\!^{-}\!} p \, V_{k-p}(t) \langle B(p,q,t) \rangle \right]. \label{eq:EOMEV}
\end{align}

Taking the expectation value of the equation of motion, making the RPA approximation, and Fourier transforming the result we obtain the much simplified equation 
\begin{align}
\omega & \langle B(k,q,\omega)\rangle = (\epsilon_ {q/2+k, \uparrow}+\epsilon_{q/2-k, \downarrow}) \langle B(k,q,\omega) \rangle  \nonumber \\
&+ [1-n^F_\uparrow(q/2+k) - n^F_\downarrow(q/2-k)]\left[{\mathcal P}^\text{ext}_q(\omega) + \int {d\!\!^{-}\!} p \, V_{k-p} \langle B(p,q,\omega)\rangle \right]. \label{eq:EOM1}
\end{align}
Reorganizing the terms in Eq.~\eqref{eq:EOM1}, we bring it to the form
\begin{align}
\langle B(k,q,\omega) \rangle \left[{\mathcal P}^\text{ext}_q(\omega) + \int {d\!\!^{-}\!} p \, V_{k-p} \langle B(p,q,\omega)\rangle \right]= \frac{1-n^F_\uparrow(q/2+k)-n^F_\downarrow(q/2-k)}{\omega - \epsilon_ {q/2+k, \uparrow}-\epsilon_{q/2-k, \downarrow}}. 
\label{eq:EOM2}
\end{align}
Integrating both sides over $k$, we obtain the integral form of the differential Eq.~\eqref{eq:EOMEV}
\begin{align}
{\mathcal P}(q,t) =\int dt' \chi_\text{pair}^{(0)} (q,t-t') \left[V(t') {\mathcal P}(q,t') +{\mathcal P}^\text{ext}_q(t')\right],
\end{align}
where $\chi_\text{pair}^{(0)}(q,t)$ is the Fourier transform of the the bare susceptibility
\begin{align}
\chi_\text{pair}^{(0)} (q,\omega)=\int {d\!\!^{-}\!} k \, \frac{1-n^F_\uparrow(q/2+k)-n^F_\downarrow(q/2-k)}{\omega - \epsilon_ {q/2+k, \uparrow}-\epsilon_{q/2-k, \downarrow}}, \label{eq:bareChi}
\end{align}
where, within the RPA approximation, $\chi_\text{pair}^{(0)}(q,t)$ is evaluated with fermions in the initial state.

A particularly simple case, is the one in which the interaction strength changes in a stepwise fashion $V(t)=V \theta(t)$. In this case, small fluctuations induced by the external field after the quench will be governed by the poles of the familiar RPA-like susceptibility
\begin{align}
\chi_\text{pair}^\text{RPA}(\omega, q) = \frac{\chi_\text{pair}^{(0)} (\omega, q)}{1- V \chi_\text{pair}^{(0)} (\omega, q)}. \label{eq:chiRPA}
\end{align}
Here, the difference between the usual RPA susceptibility and Eq.~\eqref{eq:chiRPA} is that in the former the bare susceptibility given by Eq.~\eqref{eq:bareChi} is evaluated in the equilibrium fermionic state while in the latter it is evaluated in the initial fermionic state before the quench.  If the final interaction strength falls in the pairing regime, then $\chi_\text{pair}^\text{RPA}(\omega, q)$ will have a line of poles $\omega_q=\Omega_q+i \Delta_q$ with a positive imaginary part, corresponding to the exponential growth of small fluctuations. When time reaches $t \sim 1/\Delta_\text{max}$, where $\Delta_\text{max}$ corresponds to the fastest growing mode, the amplitude of the fastest growing mode will become large. At this point in time, the various unstable modes begin to strongly interact with each other and therefore our instability analysis begins to fail as a result the initial exponential growth of the modes saturates and the topological defects freeze in. 

To get a simple physical picture of the dynamics discussed above it is useful to consider a complimentary approach that looks at the most unstable $q=0$ component of the pairing amplitude. Consider the time dependent wavefunction:
\begin{align}
|\Psi(t)\rangle = \prod_k \left(u_k(t) +v_k(t) c_{k\uparrow}^\dagger c_{-k\downarrow}^\dagger \right) 
|0 \rangle. 
\end{align}
Here $\left\{u_k(t)^2,\, v_k(t)^2\right\}=\frac{1}{2}\left(1\pm\frac{\epsilon_k}{\sqrt{\epsilon_k^2+\Delta(t)^2}}\right)$ are functions of time because the pairing gap $\Delta(t)$ is itself a function of time. The statement that energy of the collective mode in the pairing channel is $\omega_q$ is equivalent to the statement that for small values of $\Delta(t)$,  $\Delta(t)$ obeys the equation of motion
\begin{align}
\frac{d \Delta(t) }{dt} = -i \omega_{q=0} \Delta(t).
\end{align}
When $\text{Im} \, \omega_q > 0$ we get exponential growth of the pairing amplitude. It turns out that when one focuses on the $q=0$ mode only, it is possible to derive dynamics even beyond linearized approximation for $\Delta$~\cite{Levitov,Yuzbashian}. In contrast, the approach discussed above considers modes at all $q$'s but is limited to the linearized approximation.

In addition to instantaneous quenches, we can also consider ramps that occur over a finite amount of time. In this case, we can estimate the point in time at which the freezing in of topological defect occurs via a scaling argument similar to the one of Ref.~\cite{Zurek}. Suppose that the phase transition occurs at the time $t=0$. Following this time, the susceptibility computed via Eq.~\eqref{eq:chiRPA} will have a line of poles with positive imaginary frequencies. As $t$ increases the system goes deeper into the ordered phase, and the line of poles moves up to larger and larger imaginary frequencies. At each point in time, we can identify the most unstable mode and corresponding wavelength $\Delta_\text{max}(t)$ and $q_\text{max}(t)$. Approximately at the time that the inequality $t \lesssim 1/\Delta_\text{max}(t)$ is first satisfied, the fastest growing mode begins to saturate and the defects freeze in. The time dependence of $\Delta_\text{max}(t)$ can be estimated from the scaling properties of the phase transition $\Delta_\text{max}(t) \sim u^{z \nu}(t)$, where $u(t)=(V_c-V(t))/V_c$ is the time dependent distance to the phase transition, $\nu$ and $z$ are the correlation length and the dynamic critical exponents. 

At this point, we are seemingly ready to study the pairing instability. However, there is a technical difficulty that one can immediately see with the bare pairing susceptibility. The integral in Eq.~\eqref{eq:bareChi} has a UV divergence: for $k \rightarrow \infty$ the integrand becomes $-\frac{1}{2 \epsilon_k}$ and thus the integral diverges in two or more dimensions. Mathematically, the divergence originates from using a $\delta$-function (in real space) inter-atomic potential, which itself is unphysical. In the next section, we will implement a pseudopotential with a finite effective range to describe the interactions between atoms. It will turn out that scattering at low energies can be described by a universal Scattering Matrix that is independent of almost all of the details of the interatomic potential, but depends only on a few measurable scattering parameters: namely the scattering length and the effective range. We will use this knowledge to rewrite the pairing susceptibility in terms of these parameters, and thus obtain a universal description of the pairing instability.

\subsection {Dynamics of magnetization based on time dependent Hartree approximation}
Before proceeding to study the nature of the interatomic interactions, we attack the competing Stoner instability, which at first sight does not seem to suffer from a similar UV divergence. For the Stoner case, we are looking for a response to magnetization $M_q$, and therefore use the external perturbation
\begin{align}
H_\text{ext}=M^\text{ext}_q(\omega) e^{i \omega t} \sum_k c^\dagger_{k,\uparrow} c_{k+q, \uparrow} -c^\dagger_{k,\downarrow} c_{k+q, \downarrow}.
\end{align}
Within the RPA approximation, we find the susceptibility
\begin{align}
\chi_\text{FM}^\text{RPA}(\omega,q)=\frac{\chi_\text{FM}^{(0)} (\omega, q)}{1 - V \chi_\text{FM}^{(0)} (\omega, q)}, \label{eq:chiRPAFM}
\end{align}
where
\begin{align}
\chi_\text{FM}^{(0)} (\omega, q)&=\int {d\!\!^{-}\!} k \, \frac{n^F_\uparrow(q/2+k)-n^F_\downarrow(q/2-k)}{\omega - (\epsilon_ {q/2+k, \uparrow}-\epsilon_{q/2-k, \downarrow})}, \label{eq:bareChiFM}\\
&=\frac{N_0}{2}\left(1+\frac{m^2}{2 k_F q^3} \left[(\epsilon_q+\omega)^2-4\epsilon_F \epsilon_q\right] \log \frac{\epsilon_q-v_f q + \omega}{\epsilon_q+v_F q + \omega} \right. \nonumber\\
& \quad\quad\quad \quad\quad\quad  \left.+\frac{m^2}{2 k_F q^3} \left[(\epsilon_q-\omega)^2-4\epsilon_F \epsilon_q\right] \log \frac{\epsilon_q-v_f q - \omega}{\epsilon_q+v_F q - \omega}  \right),
\end{align}
and $N_0=\frac{k_F m}{2 \pi^2}$ is the density of states at the Fermi surface, and the explicit expression for the susceptibility after carrying out the ${d\!\!^{-}\!} k$ integral (in 3D) is called the Lindhard function. 

\begin{figure}
\begin{center} \includegraphics[width=10cm]{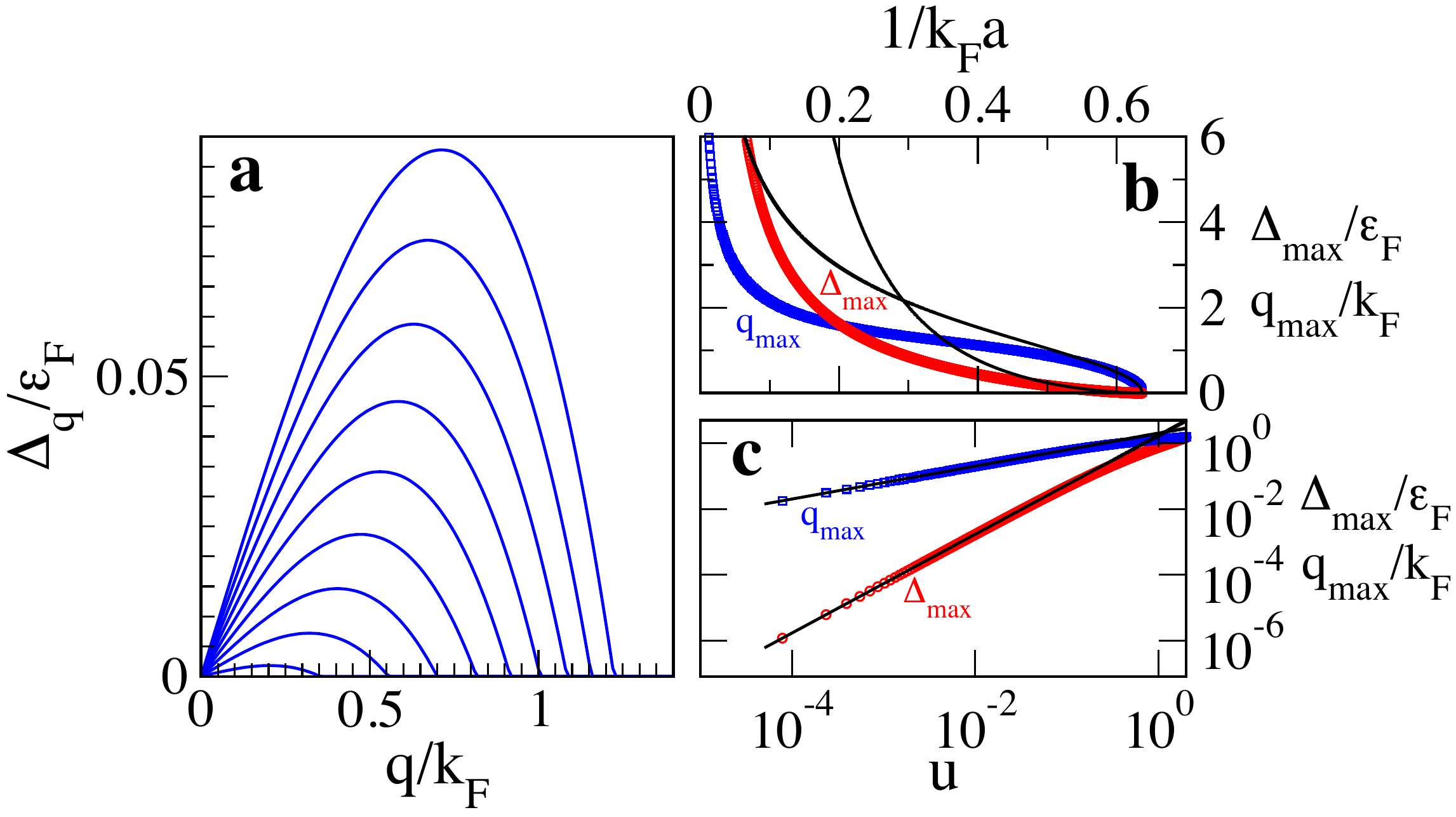} \end{center}
\caption{Properties of the unstable collective modes associated with the Stoner instability computed using $\delta$-function interactions. (a)~Growth rate $\Delta_q$ as a function of wavevector $q$ for $T=0$ and $1/k_F a=0.53$~(top line), $0.54$, $0.55$, ..., $0.63$~(bottom line).
  (b)~The most unstable wavevector $q_\text{max}$ (squares) and the
  corresponding growth rate $\Delta_\text{max}$ (circles) vs. $1/k_F a$.
  A fit to the mean-field critical theory ($\nu=1/2$, $z=3$) is shown
  with solid black lines. (c)~Details of the critical behavior of
  $q_\text{max}$ and $\Delta_\text{max}$ as a function of distance
  from the transition point $u=(1/k_Fa)_c-(1/k_Fa)$,
  $(1/k_Fa)_c=2/\pi$ on a log-log scale. }
\label{FMpoles}
\end{figure}

The susceptibility $\chi_\text{FM}^\text{RPA}(\omega,q)$ acquires purely imaginary poles $\omega_q = i \Delta_q$ for $V N_0 > 1$ and has no poles for $V N_0 < 1$. As $\Delta_q>0$, these poles correspond to unstable collective modes of the system, and therefore $V N_0=1$ corresponds to the phase transition point from Fermi liquid to Stoner Ferromagnet (at the RPA level). The growth rate of the instability $\Delta_q$ is traced out as a function of $q$ for several values of $V$ in Fig.~\ref{FMpoles}. In all cases, for small momenta $\Delta_q$ is linear in $q$. This is a reflection of the fact that magnetization is a conserved order parameter (the operator for the total spin of the system commutes with the 
Hamiltonian), therefore making large domains (small $q$) requires moving spins by large distances, which is a slow process. As $q$ increases, we are bending the emerging Ferromagnetic texture at shorter and shorter length-scales. Eventually this becomes energetically unfavorable and $\Delta_q$ bends over and decreases, becoming zero at $q=q_\text{cut}$ at which point the imaginary part of the pole of $\chi_\text{FM}^\text{RPA}(\omega,q)$ disappears. 

After a quench the fastest growing modes, i.e. those with the largest $\Delta_q$ will dominate and thus determine the size of the typical domains. In particular, we find that the for a given $u=(V-V_c)/V_c$, where $V_c$ is the interaction corresponding to the phase transition, the fastest growing modes have a $q_\text{max} \simeq 2 k_F u^{1/2}$ and a growth rate of $\Delta_\text{max} \simeq (16/3\pi) \epsilon_F u^{3/2}$. It is tempting to relate the interaction strength to the scattering length $a$ via $V=4 \pi a/m$. Implementing this temptation, we plot the growth rate as a function of inverse scattering length in Fig.~\ref{FMpoles}. Here, we start to see a problem: as the scattering length becomes larger, so does the growth rate of the Ferromagnetic instability. This is clearly a deficiency of the theory, as at unitarity (where the scattering length diverges) the only available scale is the Fermi energy scale so the distance to the transition $u$ should not appear. Physically, the deficiency lies in using the bare scattering length, which is only meaningful for low energy collisions, to describe the Stoner instability that involves all energy scales up to the Fermi energy. In the following sections we shall develop the formalism to describe collisions at all energy scales within a Fermi liquid. Afterwards, we shall come back to the case of the Stoner instability, and using a more realistic interaction potential fix the divergence at unitarity.

\section{Feshbach resonance via pseudopotentials}
\label{sec:PP}
In this section, our goal is to describe atom scattering in the vicinity of a Feshbach resonance. Our strategy is to first describe the scattering between a pair of atoms in vacuum; in the following section we shall extend this description to include Pauli-blocking to obtain a description of scattering in many-body system.

The key to controllable inter-atomic interactions in ultracold atom systems is the so called Feshbach resonance. Inter-atomic interactions in ultracold atom experiments are typically of van der Waals type and therefore intrinsically attractive (main exception to this rule are experiments with dipolar atoms and molecules). However, at low energies, interactions can appear to be repulsive. This is possible via scattering off a shallow bound state, as can be qualitatively appreciated from second order perturbation theory. A Feshbach resonance corresponds to the appearance of such a bound state. The utility of a Feshbach resonance comes from the fact that the binding energy and, therefore, the interaction strength can be tuned. The tuning is via a magnetic field as the bound state (or more appropriately the closed channel) has a slightly different magnetic moment as compared to the open channel (see~\cite{PethickSmith,Varenna} for details). 

Instead of delving into the details of atomic physics of Feshbach resonances, for pedagogical insight we present a simple model for a so called geometric resonance. In a geometric resonance, it is assumed that the inter-atomic interaction can be described by a potential that only depends on the inter-atomic distance and can be tuned directly, thus avoiding the complication of a true two channel model. 

Consider a pair of atoms with masses $m_1$ and $m_2$ interacting via the inter-atomic potential $V(r_1-r_2)$. The scattering problem is described by the Hamiltonian
\begin{align}
H=\frac{1}{2 m_1} \nabla_1^2 + \frac{1}{2 m_2} \nabla_2^2 + V(r_1-r_2). \label{eq:H2}
\end{align}
It turns out that at low energies, the details of $V$ are not important. Our main tool for describing atom scattering will be the T-matrix, which is related to the scattering amplitude in the center of mass frame via $T(E=k^2/2\mu; k,k')=-\frac{2\pi}{\mu} f(k,k')$, where $k$ and $k'$ are the relative momenta of the scattering atoms before and after the collision, $E$ is the total kinetic energy, and  $\mu^{-1}=m_1^{-1}+m_2^{-1}$ is the reduced mass of the scattering atoms. In fact, the scattering problem is universal in the sense that many inter-atomic potentials $V(r_1-r_2)$ will lead to the same form of the T-matrix for low energy scattering. Therefore, to model scattering, we can come up with any suitable potential that produces the T-matrix we want. This type of model potential is usually called a pseudopotential.

\begin{figure}
\begin{center}
\includegraphics[scale=0.23]{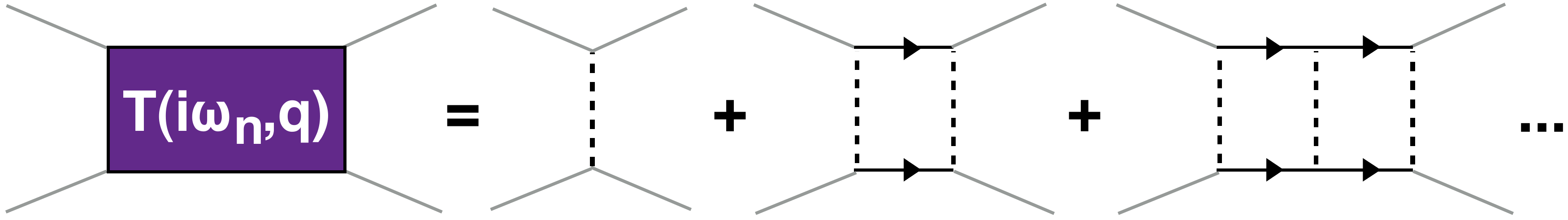}
\vskip1cm
\includegraphics[scale=0.23]{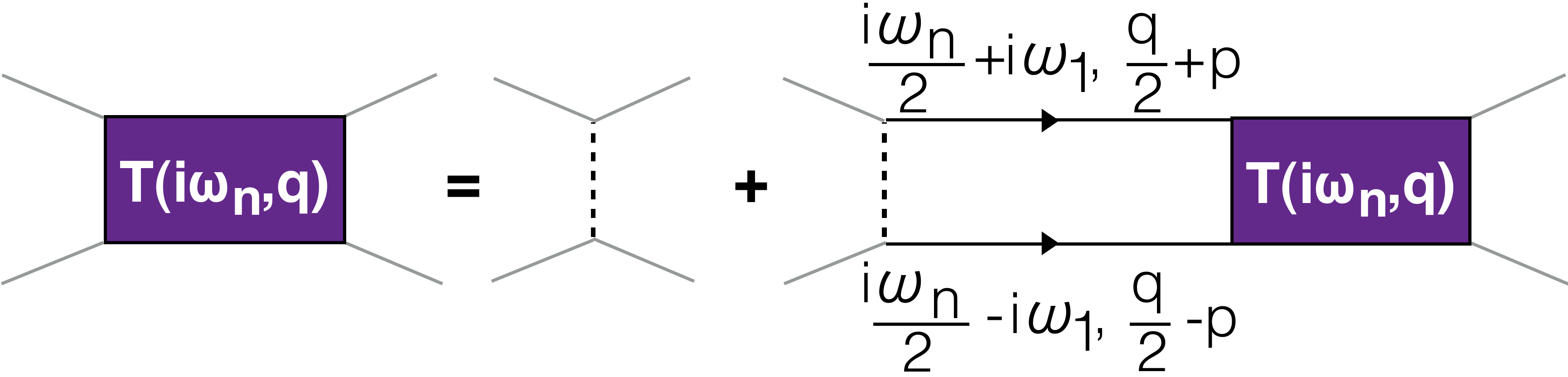}
\end{center}
\caption{Graphic representation of the T-matrix as a summation of Ladder diagrams (top) and the corresponding Lippmann-Schwinger equation (bottom). Purple squares represent the T-matrix, solid black lines fermionic atom propagators, dotted black lines inter-atom interactions, and solid grey lines external fermionic atom legs. The external legs are shown for clarity, and indicate how one would go about connecting the T-matrix to fermionic lines.}
\label{fig:Ladder}
\end{figure}

The poles of the T-matrix correspond to two atom bound states. Near a Feshbach resonance, there must be a bound state having a spatial extent of the scattering length $a$, which corresponds to a binding energy of $E_b \approx 1/2\mu a^2$. Therefore the T-matrix must have the form 
\begin{align}
T(E)=-\frac{2\pi}{\mu}\left(-\frac{1}{a}-i\sqrt{2 \mu E} + r_e \mu E + O(E^2)\right)^{-1},
\label{eq:Tmatrix}
\end{align}
where $a$ is the scattering length and $E$ is the kinetic energy of the two scattering atoms in the center of mass frame. The effective range $r_e$ is the first correction of the binding energy due to the shape of the inter-atomic potential and roughly corresponds to its spatial extent. Here, we have specialized to the case of s-wave scattering and thus the T-matrix has no angular dependence.

The T-matrix may be obtained directly by soling the Schrodinger equation, Eq.~\eqref{eq:H2}.  An alternative, and more instructive approach, is to obtain the T-matrix via a resummation of particle-particle ladder diagrams (see Fig.~\ref{fig:Ladder}) which yields the Lippmann-Schwinger equation
\begin{align}
T(i \omega_n, q; k,k')&=V(k-k') +\sum_{i \omega_1} \int {d\!\!^{-}\!} k_1 V(k-k_1) G_\uparrow(\frac{i \omega_n}{2}+i \omega_1, \frac{q}{2}+k_1) \nonumber \\
&\quad\quad\quad\quad\quad\quad\quad\quad\quad G_\downarrow(\frac{i \omega_n}{2}-i \omega_1, \frac{q}{2}-k_1) T(i \omega_n, q; k_1, k'),
\label{eq:LS}
\end{align}
where $G_\sigma(i \omega_n, q)=(i \omega_n - k^2/2m)^{-1}$ is the free fermion Green function, and we have added the center of mass momentum $q$ to the labels of the T-matrix. Since the physical T-matrix does not depend on the relative momenta differences, we shall drop them from our notation.

Many different pseudopotentials will result in the same T-matrix Eq.~\eqref{eq:Tmatrix}. In fact, many different pseudopotentials are in common use, e.g. box potentials, Gaussian potential, hard sphere potentials, and regularized $\delta$-function potentials, see e.g.~\cite{PethickSmith,Pilati,Trivedi}. Here, following Ref.~\cite{Phillips}, we shall use a slightly less common form of the pseudopotential 
\begin{align}
\lambda(k_1, k_2) = \sum_{i,j=0}^1 \lambda_{ij} k_1^{2 i} k_2^{2 j},
\end{align}
where the matrix $\lambda_{ij}$
\begin{align}
\lambda_{ij}=\left(
\begin{array}{cc}
C & C_2\\
C_2 & 0
\end{array}
\right).
\end{align}
This form has two tuning parameters $C$ and $C_2$ which are needed to match both the scattering length and effective range which appears in the T-matrix, allowing for the flexibility to describe both wide and narrow resonances. For the special case of a wide resonance, where $r_e \approx 0$, we can drop $C_2$, thus eliminating the matrix structure and reducing the description to that of Ref.~\cite{PethickSmith}. The main advantage of this form of the pseudopotential is that it is separable, thus considerably simplifying the Lippmann-Schwinger Eq.~\eqref{eq:LS}. We can now write the T-matrix in the same form as the pseudopotential
\begin{align}
T(E, k_1, k_2)=\sum_{i,j=0}^1 \tau_{ij}(E) k_1^{2 i} k_2^{2 j},
\end{align}
where the $\tau(E)$ matrix is to be determined. Due to the translational invariance of the problem,  we shall work in the center of mass frame. The T-matrix in a moving frame is related to the T-matrix in the center of mass frame via $T(E,q)=T(E-q^2/2(m_1+m_2),0)$. From here on, we shall take the step of setting $m_1=m_2=m=2\mu$ to simplify the notation. In matrix form the Lippmann-Schwinger equation becomes
\begin{align}
\tau(E)=\lambda + \lambda I(E) \tau(E) \label{eq:LSM}
\end{align}
where
\begin{align}
I(E)=
\left(
\begin{array}{cc}
\int \frac{d^3k}{(2 \pi)^3} \frac{1}{E^+-k^2/2\mu} & \int \frac{d^3k}{(2 \pi)^3} \frac{k^2}{E^+-k^2/2\mu}\\
\int \frac{d^3k}{(2 \pi)^3} \frac{k^2}{E^+-k^2/2\mu} & \int \frac{d^3k}{(2 \pi)^3} \frac{k^4}{E^+-k^2/2\mu}
\end{array}
\right). \label{eq:IofE}
\end{align}
The integrals in \eqref{eq:IofE} come with an upper cut-off $\Lambda$, and $E^+$ stands for $E+i \delta$. We see that the divergence that appears in the bare pairing susceptibility, Eq.~\eqref{eq:bareChi}, is of exactly the same type as the $11$ component of the $I(E)$ matrix, suggesting that the two are related. 

Having obtained a relation between the pseudopotential and the physically observable T-matrix, we can plug the pseudopotential into the many-body problem to obtain an effective interaction parameters $C$ and $C_2$ in terms of the scattering length, effective range, and cut-off. Although the parameters $C$ and $C_2$ depend on the cut-off, as we shall demonstrate the pairing susceptibility is be independent of it. 

\section{Application to pairing susceptibility}
\label{sec:Apply}
The pairing susceptibility, that we originally obtained using the equation of motion approach, can also be obtained diagrammatically. Explicitly, the susceptibility is related to the two particle propagator, i.e. the Cooperon, which is the many body version of the T-matrix. At the RPA level, the Cooperon $C(i\omega_n, q)$ corresponds to the solution of the Lippmann-Schwinger equation~\ref{eq:LS}, with free Green functions replaced by Green functions of fermions in a Fermi sea $G_\sigma(i \omega_n, q)=(i\omega_n-k^2/2m+\epsilon_F)^{-1}$, where $\epsilon_F$ is the Fermi energy. Thus, at the RPA level the difference between the Cooperon and the T-matrix is that the Cooperon takes into account Fermi blocking so that scattering only occurs on top of the Fermi sea. 

\begin{figure}
\begin{center}
\includegraphics[scale=0.3]{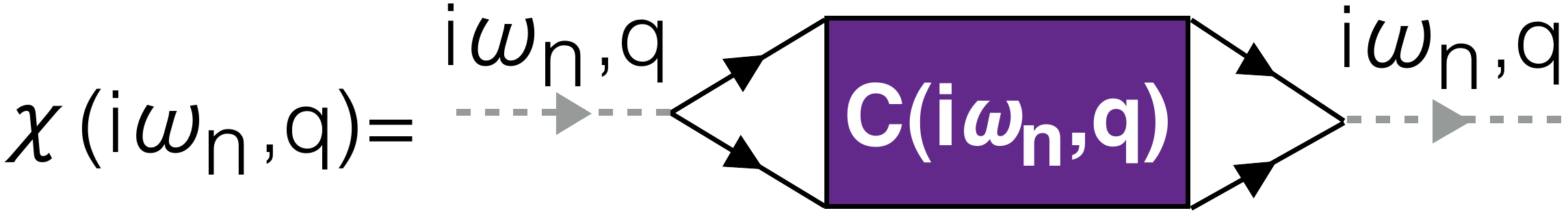}
\end{center}
\caption{Diagrammatic representation of the relation between the Cooperon (many-body T-matrix) and the pairing susceptibility. Solid black lines represent fermionic atom propagators, purple square represents the Cooperon, and dotted grey lines represent external source of the pairing field.}
\label{fig:Chi}
\end{figure}

The relationship between $C(i\omega_n, q)$ and $\chi(i\omega_n, q)$ is shown schematically in Fig.~\ref{fig:Chi}.  Explicitly, the relationship is
\begin{align}
\chi(i \omega_n, q)& =\chi^{(0)} (i \omega_n, q)\nonumber\\
&+\sum_{i \omega_1, i \omega_2} \int {d\!\!^{-}\!} k_1\, {d\!\!^{-}\!} k_2 \, G_\uparrow(\frac{i \omega_n}{2}+i\omega_1, \frac{q}{2}+k_1) G_\downarrow(\frac{i \omega_n}{2}-i\omega_1, \frac{q}{2}-k_1) \nonumber \\
&\quad \times C(i \omega_n, q; k_1, k_2) G_\uparrow(\frac{i \omega_n}{2}+i\omega_1, \frac{q}{2}+k_1) G_\downarrow(\frac{i \omega_n}{2}-i\omega_1, \frac{q}{2}-k_1).
\end{align}
Using this relation, it can be shown we can recover Eq.~\eqref{eq:chiRPA} for the susceptibility. At this point, we remark that the poles of the Cooperon and the pairing susceptibility match. Therefore, to find the unstable collective modes it is sufficient to look at the poles of the Cooperon.

\begin{figure}
~\vtop{\vskip-1ex \hbox{\includegraphics[width=6cm]{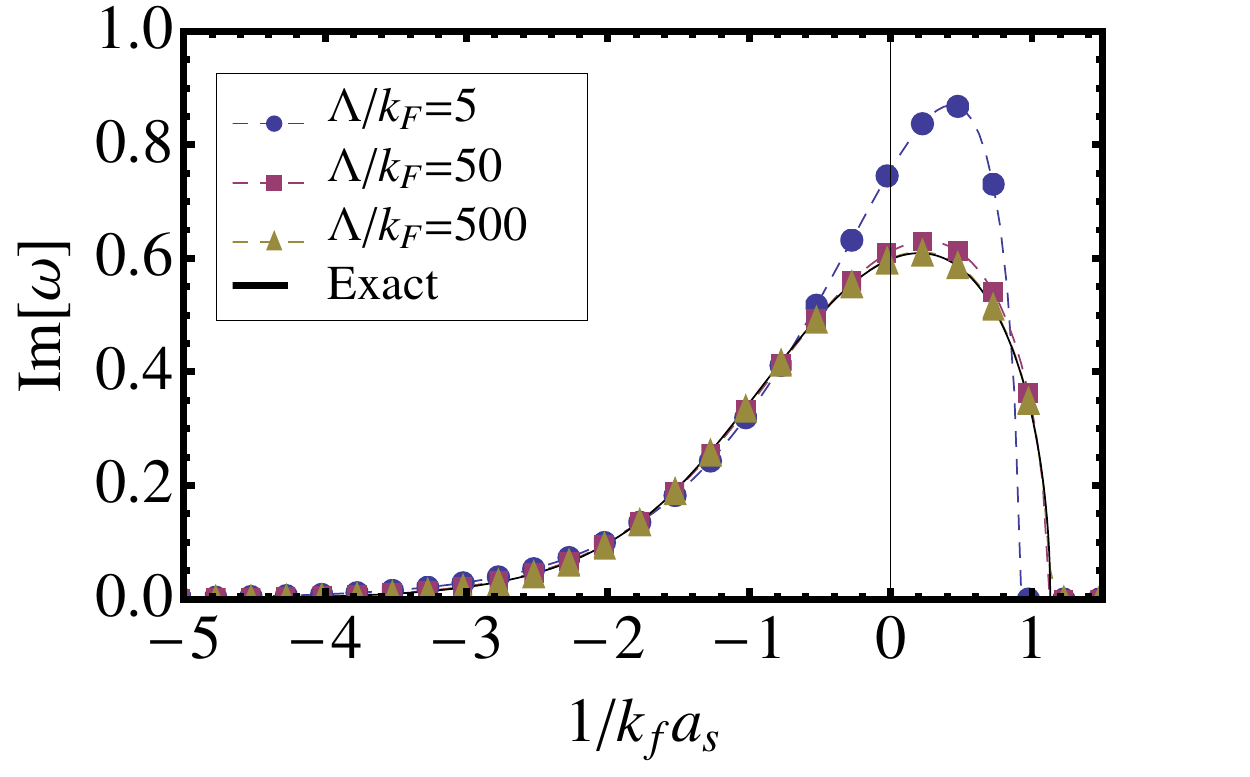}}}
\hspace{-6cm}{\bf a} \hspace{6cm}
\hskip0.1cm
\vtop{\vskip-1ex \hbox{\includegraphics[width=6cm]{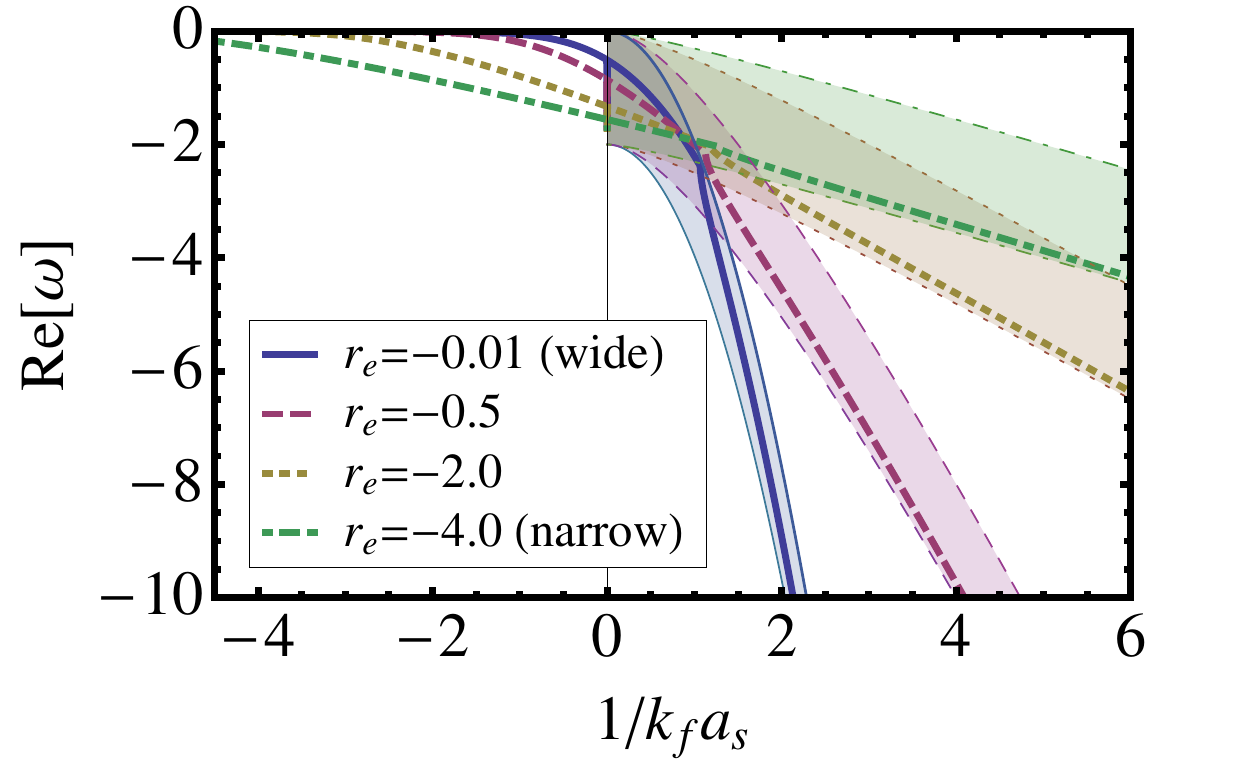}}}
\hspace{-6cm}{\bf b} \hspace{6cm}~\\
~\vtop{\vskip-1ex \hbox{\includegraphics[width=6cm]{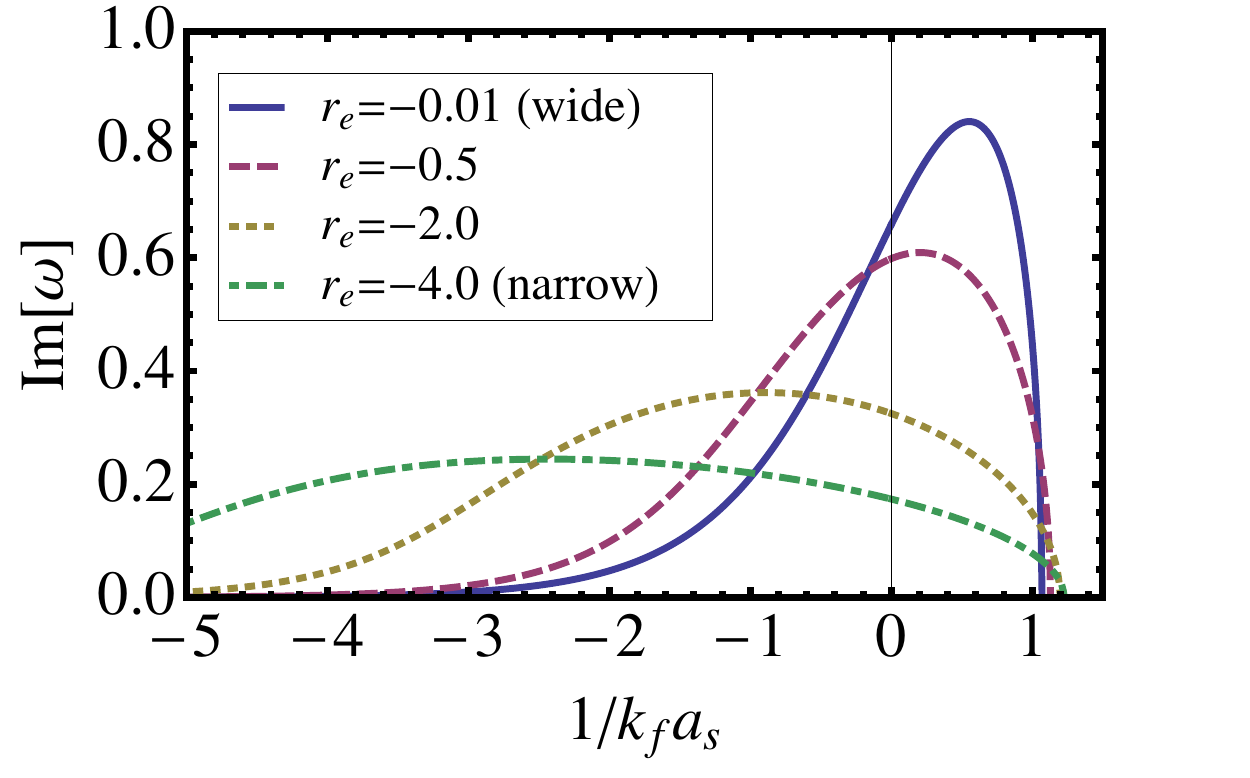}}}
\hspace{-6cm}{\bf c} \hspace{6cm}
\hskip0.1cm
\vtop{\vskip-1ex \hbox{\includegraphics[width=6cm]{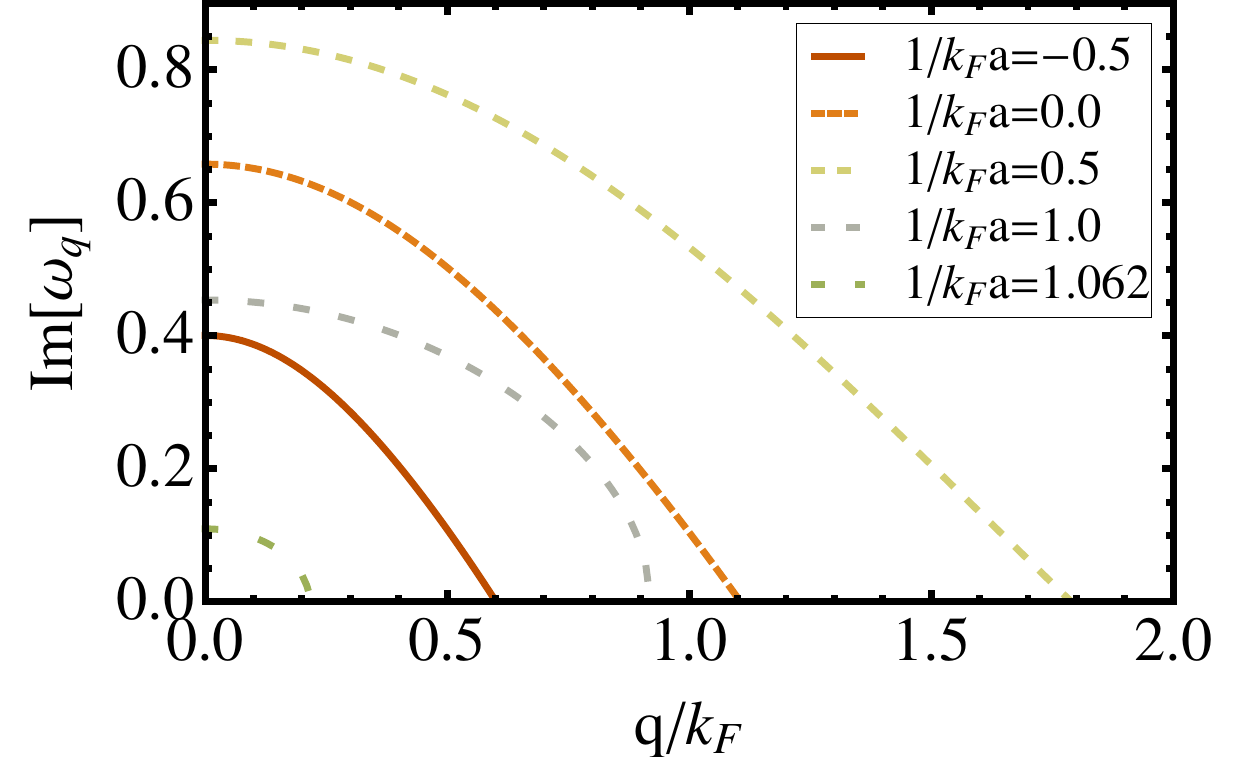}}}
\hspace{-6cm}{\bf d} \hspace{6cm}~
\caption{(a) Imaginary part of the Cooperon pole trajectory as a function inverse scattering length for various values of the cut-off $\Lambda$ and $r_e=-0.5$. As $\Lambda$ increases the pole trajectory approaches its asymptotic value. (b) Real part of the Cooperon pole trajectory as a function inverse scattering length for various values of the effective range $r_e$. 
The trajectory of the corresponding vacuum binding energies (T-matrix poles) and vacuum binding energies shifted down by the twice the Fermi energy (T-matrix poles - $2\epsilon_F$) are indicated by the tops and bottoms of the shaded corridors. In the strong interaction regime (small positive scattering length) the Cooperon pole approaches the bottom of the corridor, i.e. the T-matrix pole shifted down by twice the Fermi energy.  
(c) Imaginary part of the Cooperon pole trajectory (pairing rate) as a function inverse scattering length for various values of the effective range $r_e$. (d) Imaginary part of the Cooperon pole trajectory (pairing rate) as a function of momentum for various scattering lengths, and a wide resonance ($r_e=-0.01$). The plot indicates that the most unstable pairing mode is always at $q=0$.}
\label{fig:CooperonPoles}
\end{figure}

In matrix form, the analog of the Lippmann-Schwinger equation~\eqref{eq:LSM} for the T-matrix is the Lippmann-Schwinger equation for the Cooperon (using RPA)
\begin{align}
{\mathcal C}(E,q)=\lambda + \lambda \tilde{I}(E,q) {\mathcal C}(E,q)
\end{align}
where $\tilde{I}(E,q)$ 
\begin{align}
\tilde{I}(E)=
\left(
\begin{array}{cc}
\int \frac{d^3k}{(2 \pi)^3} \frac{1-2n_F(k)}{E^+-k^2/2\mu + 2 \epsilon_F-q^2/4m} & \int \frac{d^3k}{(2 \pi)^3} \frac{k^2(1-2n_F(k))}{E^+-k^2/2\mu + 2 \epsilon_F-q^2/4m}\\
\int \frac{d^3k}{(2 \pi)^3} \frac{k^2(1-2n_F(k))}{E^+-k^2/2\mu + 2 \epsilon_F-q^2/4m} & \int \frac{d^3k}{(2 \pi)^3} \frac{k^4(1-2n_F(k))}{E^+-k^2/2\mu + 2 \epsilon_F-q^2/4m}
\end{array}
\right). \label{eq:ITofE}
\end{align}
At this point we can use the relationship between interaction matrix $\lambda$ and the physical parameters $a$ and $r_e$ to compute the Cooperon and find its poles. The resulting plot of the poles is shown in Fig.~\ref{fig:CooperonPoles}. To understand the pole structure, we first comment on the effect of the cut-off. As the momentum cut-off $\Lambda$ is increased, the parameters $C$ and $C_2$ that appear in $\lambda$ matrix change. However, as can be clearly seen in Fig.~\ref{fig:CooperonPoles}a, the Cooperon poles converge to their asymptotic values once $\Lambda$ exceeds the Fermi-momentum by a factor of $\sim 100$.

Instead of dealing with the cut-off, we can directly cancel the divergences in the Cooperon by comparing it with the T-matrix. Comparing Eq.~\ref{eq:IofE} and Eq.~\ref{eq:ITofE}, we find that denominators of $I(E+2\epsilon_F-q^2/4m)$ and $\tilde{I}(E)$ match, thus the divergences of $C(E,q)$ and $T(E+2\epsilon_F-q^2/4m,q=0)$ cancel. Using this fact, we add and subtract the T-matrix to the Cooperon Lippmann-Schwinger equation
\begin{align}
{\mathcal C}^{-1}(E,q)&=\lambda^{-1}-\tilde{I}(E)+\tau^{-1}(E+2\epsilon_F-q^2/4m)-\tau^{-1}(E+2\epsilon_F-q^2/4m).
\end{align}
Carefully inverting this equation shows that only the $11$ component is non-zero, and we obtain the expression
\begin{align}
C(E,q)=&\left[ \frac{\mu}{2\pi} \left(\frac{1}{a}+i\sqrt{2\mu\left(E+2\epsilon_F-\frac{q^2}{4m}\right)}-r_e \mu \left(E+2\epsilon_F-\frac{q^2}{4m}\right) \right)\right. \nonumber \\
&\quad\quad\quad\quad\quad\quad\quad\quad\quad \left.+\int {d\!\!^{-}\!} k \frac{n_F(k+q/2) + n_F(k-q/2)}{E+2\epsilon_F-k^2/m-q^2/4m}\right]^{-1}.
\end{align}
In Fig.~\ref{fig:CooperonPoles}a, we compare the pole structure obtained by gradually increasing the cut-off $\Lambda$ with the asymptotic pole structure obtained from the above expression. In doing so we verify that the asymptotic expression is indeed correct and corresponds to $\Lambda \rightarrow \infty$. 

Having understood how to regularize the divergence in the Lippmann-Schwinger equation, we come back to the question of understanding the pole structure.
We begin our analysis with the T-matrix in vacuum. For each value of
$a$, $T(E,q)$ has a line of poles on the BEC
side located at $E=\omega_q+i\Delta_q=-1/ma^2+m q^2/4$,
corresponding to the binding energy of a Feshbach molecule with center
of mass momentum $q$. As a consequence of energy and momentum
conservation the pole frequency is real, indicating that a two-body process in vacuum 
cannot produce a Feshbach molecule. 

The Cooperon is a natural extension of the two particle scattering amplitude, i.e. the T-matrix, to a systems with a finite density of atoms. The presence of the Fermi-sea shifts the poles of the Cooperon relative to the T-matrix in two ways: (1) in the range $-\infty<1/k_F a\lesssim 1.1$ the Cooperon pole acquires a positive imaginary part $\Delta_q$ that corresponds to the growth rate of the pairing instability (see Fig.~\ref{fig:CooperonPoles}a); and (2) the real part of the pole $\Omega_q$, which would correspond to the binding energy of a pair in the absence of an imaginary part, uniformly shifts down (see Fig.~\ref{fig:CooperonPoles}b). 

The shift of the pole into the complex plane is quite surprising, especially on the BEC side of the Feshbach resonance where the interactions are repulsive, such a shift corresponds to a finite rate of molecule formation. Typically one assumes that a two particle collision can not lead to the formation of a molecule as energy and momentum conservation laws can not be satisfied simultaneously, and hence earlier analysis focused on considering  at least a three body collision~\cite{Shlyapnikov}. Hence, one would think that by analyzing poles of the Cooperon, which seems to describe two particle collisions, we cannot get a pole with a finite imaginary part.  An important difference of our system is that we are considering a many-body system. So even though the Cooperon can be understood as effectively a two particle scattering amplitude, it describes a scattering event taking place in the presence of a filled Fermi sea. The Pauli principle plays a role of the ``third body".  That is the energy-momentum restrictions on molecule (or more precisely Cooper pair) formation are lifted, as the pair forms above the Fermi-sea, the excess energy can be absorbed by the two holes that are left behind under the Fermi-sea.  This process is schematically represented in the inset of Fig.~\ref{fig:composite}. Within our approximation there is a sharp cut-off of the molecule formation rate on the BEC side when distance between particles becomes much larger than the scattering length and the Pauli principle becomes ineffective. In reality, we expect that the molecule formation rate does not go to zero completely but becomes determined by much slower three body processes discussed earlier in Refs.~\cite{Petrov,other3body1,other3body2,other3body3}.

The uniform shift down of the real part of the pole $\Omega_q$ is likewise a result of Pauli
blocking~\cite{AGD}, and indicates an appearance of a paired state
on the BCS side as well as stronger binding of the pairs on the BEC
side.  Deep on the BEC side, the Feshbach molecule becomes deeply bound
and therefore very small in real space. As a result the molecule becomes 
extended far beyond $k_F$ in momentum space and Pauli blocking becomes less relevant.
Consequently, see Fig.~\ref{fig:CooperonPoles}b, the Cooperon pole (solid lines) approaches the T-matrix pole (shifted by $2\epsilon_F$, dotted lines) deep on the BEC side.

Coming back to the imaginary part of the pole, as depicted in
Fig.~\ref{fig:CooperonPoles}c, $\Delta_{q=0}$
increases exponentially as one approaches the Feshbach resonance from the BCS
side, {\it i.e.} the growth rate of the BCS pairing in a wide resonance is equal to the
BCS gap at equilibrium $\Delta_{q=0} \approx 8 \epsilon_F e^{\pi/2
  k_F a-2}$~\cite{AGD}. For a wide resonance,  the growth rate
continues to increase on the BEC side, reaching a maximum at $k_F a \approx 2$, and
finally decreasing to zero at $k_F a \approx 1.1$, at which point the Fermi
sea can no longer absorb the energy of the Feshbach molecule in a
two-body process. Deeper in the BEC regime pairing takes place via the
more conventional three-body process and would round the pairing
instability curve near $k_F a \approx 1.1$ in Fig.~\ref{fig:CooperonPoles}c. 
As we go from a wide resonance $r_e \ll 1/k_F$, to a more narrow resonance 
$r_e \approx 1/k_F$, the maximum in the paring rate decreases and shifts 
to the BEC side (see Fig.~\ref{fig:CooperonPoles}c).

We comment that pairing at finite $q$ is always slower than at $q=0$, with $\Delta_q$ monotonically decreasing to zero at $q=q_\text{cut}$ (see Fig.~\ref{fig:CooperonPoles}d). Throughout a wide resonance the approximation
$q_\text{cut} \approx (\sqrt{3/2}) (\Delta_{q=0}/\epsilon_F) k_F$
works reasonably well except in the vicinity of $k_f a \sim 2$ where
$q_\text{cut}$ reaches the maximal value for a two-body process of $2 k_f$.

\section{More on Stoner instability}
\label{sec:StonerWithCooperon}
\begin{figure}
\begin{center}
\includegraphics[scale=0.8]{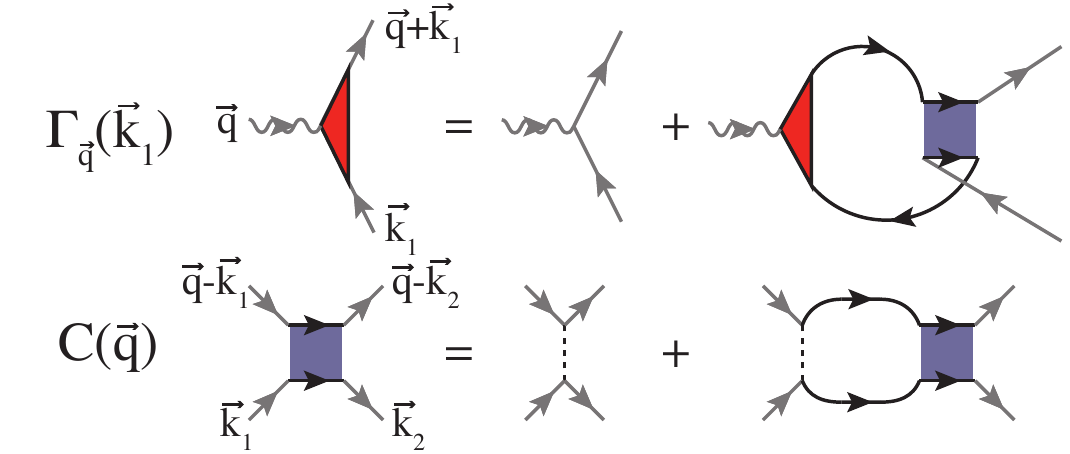}
\end{center}
\caption{
Diagrammatic representation of the vertex function $\Gamma_{\vec{q}}(\vec{k}_1)$ 
that appears in the Ferromagnetic susceptibility with the bare two particle interaction $V$ replaced by the Cooperon $C(\vec{q}_1)$, which is a momentum and frequency dependent interaction. Curly lines represent external sources of spin flips, solid lines -- fermions, dashed lines -- interactions, gray lines -- external legs. Figure is reprinted from Ref.~\protect\cite{us}
}
\label{fig:GammaDiag}
\end{figure}

In section~\ref{sec:LR}, we left the story of the Ferromagnetic instability at the unphysical divergence of its growth rate in the unitary regime. In this section, we use the knowledge gained in the previous two section to fix this divergence. The reason for the divergence, lies in the description of inter-particle interactions near unitarity. Even without a Fermi-sea, from the form of the T-matrix \eqref{eq:Tmatrix}, we see that the interactions are strongly frequency dependent $T(\omega) \approx 4 \sqrt{2} \pi / i m \sqrt{\omega}$. Only at very low energies does the expression $T(\omega) \approx 4 \pi a/m$, that we have used for the interaction strength in the calculation of the Stoner instability makes sense. 

To proceed, we replace the interatomic interaction by the Cooperon. In fact, this program has been implemented before in the context of the fermionic Hubbard model, see e.g. Ref~\cite{HubbardDiags}. Instead of computing the Ferromagnetic susceptibility directly, it is advantageous to compute the vertex function as indicated in Fig.~\ref{fig:GammaDiag}. The susceptibility is related to the vertex function via
\begin{align}
\chi_\text{FM} (\vec{q})=\int d\vec{k}_1 \, G(\vec{q}+\vec{k}_1) G(\vec{k}_1) \, \Gamma_{\vec{q}} (\vec{k}_1),
\end{align}
where we have switched to the notation $\vec{q}=(\omega,q)$ in order to save space. The poles of the susceptibility arise due vertex function and not the two Green functions in the above expression. Therefore, to find the poles of the susceptibility it is sufficient to find the poles of the vertex function. However, this is a rather complicated task, and necessarily involves some approximations. Instead of going through the details, which are presented elsewhere~\cite{us}, here we comment on the physics of the results and the differences between using the Cooperon and the $\delta$-function interactions.

\begin{figure}
\begin{center}
\includegraphics[width=10cm]{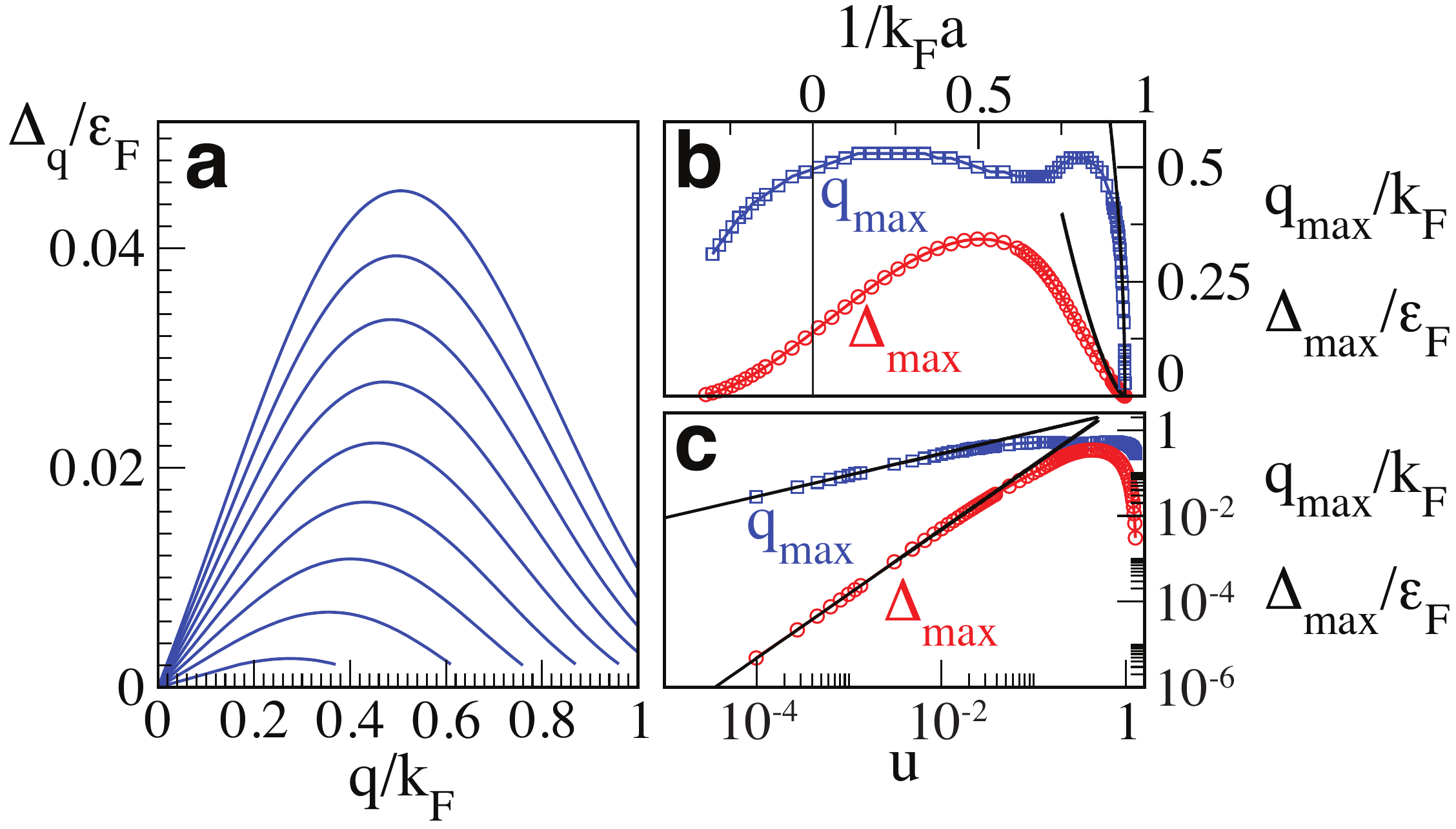}
\end{center}
\caption{Properties of the unstable collective modes associated with the Stoner instability computed using Cooperon interactions, as opposed to $\delta$-function interactions of Fig.~\ref{FMpoles}. 
(a)~Near the transition, $\Delta_q$ curves look similar to the $\delta$-function case, except the critical point has changed so correspondingly the curves were computed for $1/k_F a=0.85$~(top line), $0.86$, $0.87$, ..., $0.93$~(bottom line). (b)~$q_\text{max}$ (squares) and $\Delta_\text{max}$ (triangles) vs. $1/k_F a$.
  A fit to the mean-field critical theory ($\nu=1/2$, $z=3$) is shown with solid black lines. Note that the unphysical divergence at the Feshbach resonance, that was present for $\delta$-function interactions, has disappeared. Further, the instability continues to the attractive side of the resonance. (c)~Details of the critical behavior of  $q_\text{max}$ and $\Delta_\text{max}$ as a function of distance
  from the transition point $u=(1/k_Fa)_c-(1/k_Fa)$,
  $(1/k_Fa)_c\approx0.94$.  Figure reprinted from Ref.~\protect\cite{us}
  }
\label{fig:Stoner2}
\end{figure}

The most significant difference is the disappearance of the divergence of the instability rate near the resonance. Moreover, not only does the rate of the Stoner instability become finite everywhere, the instability persists on the BCS (attractive) side. The singularity of the Stoner instability rate was related to the singularity of the scattering amplitude of two particles in vacuum at zero energy. When analyzing a many-body system we need to integrate over energies of the order of the Fermi energy. Since for any finite energy, there is no singularity in the scattering amplitude, this leads to a suppression of the Stoner instability rate. Furthermore, the Pauli blocking by the Fermi sea no longer allows to make an ``ideal" Feshbach molecules. Instead, these molecules are restricted to occupy states outside of the Fermi momentum, which in fact enhances their binding energy. The strongest Stoner instability corresponds to the $k_Fa$ at which the scattering at typical energies is strongest. The typical energy scale is the Fermi energy $\epsilon_F$, and scattering is strongest when the bound state energy correspond to the typical energy scale. Thus the Stoner instability is strongest not when the bound state disappears in vacuum but in the vicinity of the point where the bound state energy plotted in Fig.~\ref{fig:CooperonPoles}b ``crosses" $\epsilon_F$. This crossing occurs on the BEC side of the resonance. As a result, the Feshbach resonance in vacuum is not reflected in any singular structure in the presence of a Fermi sea. Instead, the $\Delta_\text{max}$ has a maximum on the BEC side, and smoothly decreases to zero on the BCS side.

Another minor difference for the Stoner instability between the $\delta$-function and the Cooperon interactions, is that the location of the phase transition point, which shifts from $(1/k_F a)_c = 2/\pi$ for $\delta$-function interactions to $(1/k_F a)_c \approx 0.94$ for Cooperon interactions. The shift of the phase transition point is again associated with the form of the interactions at higher energies. However, the behavior of $\Delta_q$'s in the vicinity of the phase transition is very similar for both cases as can be seen by comparing Figs.~\ref{FMpoles} and \ref{fig:Stoner2}. Indeed, the power laws describing the behavior of $q_\text{max}$ and $\Delta_\text{max}$ remain the same. 

Finally, we comment on the nature of approximations that we make in computing the ferromagnetic susceptibility. In resumming diagrams indicated in Fig.~\ref{fig:GammaDiag}, we did not resum all interaction terms. We resummed only the most divergent contributions as the scattering length was going to infinity. In particular, the diagrams we resum correspond to a very simple time dependent Hartree type analysis in which we neglected changes in the Fermi occupation numbers and screening of interactions. Our justification was that we are interested in short time dynamics when such processes can be neglected. However it is possible that we have a hierarchy of time scales. Dynamics of order parameters may be slow due to the usual critical slowing down near the quantum phase transition. On the other hand  there may be fast processes on the timescale of $\epsilon_F^{-1}$, which we may not be able to take into account. Although, in the vicinity of the Feshbach resonance and away from the phase transition, the order parameter dynamics also occurs on time scales of  $\epsilon_F^{-1}$, indicating that there is no timescale separation. A more careful Keldysh type analysis of nonequilibrium dynamics, may be required.

\section{Discussion}
\label{sec:Discussion}
\begin{figure}
\begin{center}
\includegraphics[width=8cm]{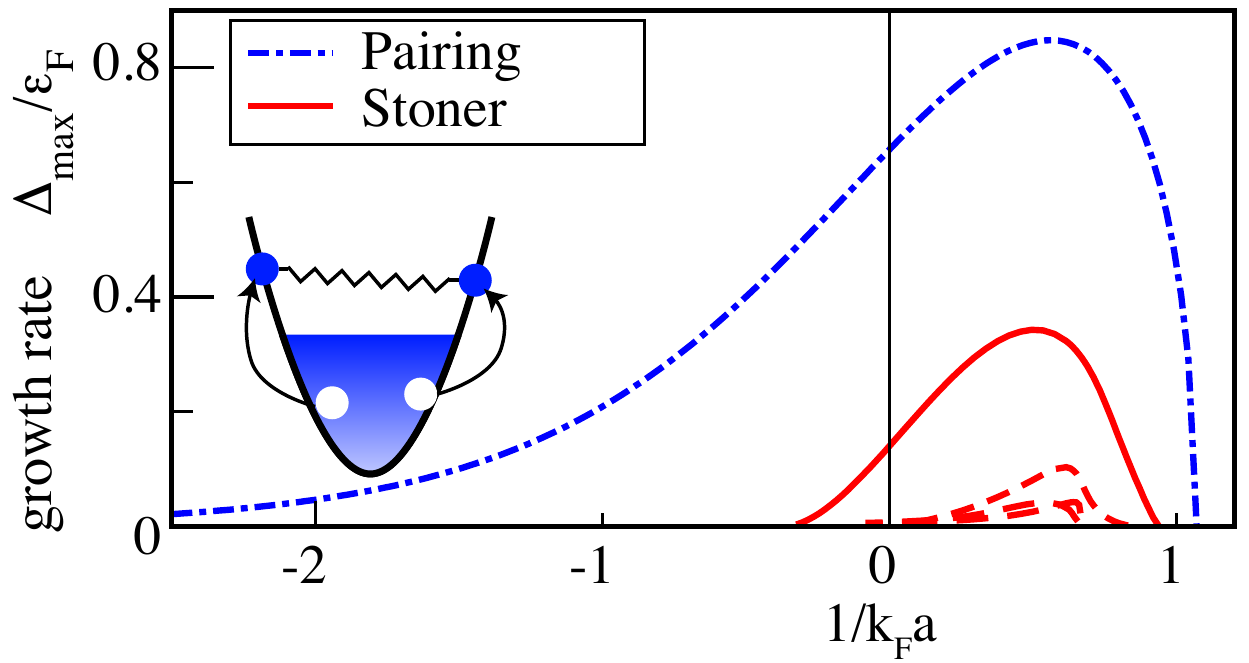}
\end{center}
\caption{Comparison of the growth rates of the pairing instability (dashed-dotted blue line) and the Stoner instability (solid red line) across the Feshbach resonance as a function of $1/k_F a$. The Stoner instability can occur in several ``angular momentum" channels, the subdominant channels are traced by dashed red lines. Finally, we also plot the growth rate for the Stoner instability obtained with RPA and $\delta$-function interactions that shows unphysical divergence near the resonance. 
  {\bf Inset}: Schematic diagram of the pair creation process
  showing the binding energy (spring) being absorbed by the Fermi
  sea (arrows). The figure is reprinted from Ref.~\protect\cite{us}.
  }
\label{fig:composite}
\end{figure}

We summarize the results obtained thus far in Fig.~\ref{fig:composite}. We find that after a quench from the weakly interacting regime to the vicinity of the Feshbach the pairing and the Stoner instabilities compete with each other on both sides of the resonance. The growth rate associated with the pairing instability is always larger than that of the Stoner instability indicating that a paired state is thus the favored outcome. In treating both instabilities, we found that it is important to describe the interactions carefully. 

Finally, we comment on the interpretation of the MIT experimental observations of Ref.~\cite{Ketterle} in light of the competition between pairing and ferromagnetism. Following the ramp to the strongly interacting regime, the MIT group let the system evolve for some time before performing their measurements. The most striking results were obtained for the atom loss measurement, which showed that as the scattering length is increased the atom loss first increases, and then suddenly starts to decreases for $k_F a \gtrsim 2$. The loss rate was measured by rapidly ramping down the magnetic field at the end of the experiment and thus projecting weakly bound "Feshbach molecules" onto strongly bound molecules away from the resonance. Thus the rapid decrease of the atom loss can have two interpretations: (1) the formation of ferromagnetic domains prevented atomic collisions (which occur only between fermions of different species) and thus resulted in a decrease of atom loss rate, or (2) the atom losses are caused by pair formation and the maximum of the pairing rate near $k_F a = 2$ corresponds to the maximum in atom loss rate. Interpretation (1) has been studied in a series of theoretical papers~\cite{Gareth} and has been shown to be reasonably consistent with experimental observations. Since we find that the pairing instability always dominates over the Stoner instability we are forced to conclude that scenario (2), which is also consistent with experimental observations, is more likely.  

In addition to the atom loss rate the MIT group, likewise, measured the changes of the cloud size and the average kinetic energy. Indeed, mean field theory calculations for the Stoner transition in a trap~\cite{Arun} show similar trends to those found by the MIT experiments. However, the pairing transition is quite similar to the Stoner transition in the sense that the atoms gain potential energy at the cost of kinetic energy. Thus, we expect that the trends for cloud size and kinetic energy would be similar for the two transitions.

\section{Concluding remarks}
\label{sec:Conclude}
The notion of collective modes is important in understanding not only equilibrium physics but also dynamics. As an example, we have investigated the role of the pairing and the ferromagnetic modes in quenches across the Feshbach resonance where these unstable modes directly compete with each other in real time. 

We acknowledge our collaborators on the original work~\cite{us}, which formed
the basis of these lecture notes: M. Babadi, L. Pollet, R. Sensarma, N. Zinner, and M. Zwierlein. We also acknowledge stimulating discussions with A. Georges, W. Ketterle,
D. Huse, G. Shlyapnikov. This work was supported by the
Army Research Office with funding from the DARPA OLE
program, Harvard-MIT CUA, NSF Grant No. DMR-07-05472, 
AFOSR Quantum Simulation MURI, AFOSR MURI on Ultracold Molecules, 
the ARO-MURI on Atomtronics.

\bibliographystyle{OUPnamed_notitle}
\bibliography{fullBib}

\end{document}